\documentclass[12pt,aps,nofootinbib]{revtex4}

\usepackage{amsmath,amssymb,bm}
\usepackage[dvips]{graphicx}
\usepackage{hyperref}
\usepackage{yfonts}
\usepackage{color}

\newcommand{\beq}{\begin{eqnarray}}
\newcommand{\eeq}{\end{eqnarray}}
\renewcommand\d{\partial}

\begin{document}

\title{Photonic chiral vortical effect}

\author{Naoki Yamamoto}
\affiliation{Department of Physics, Keio University, Yokohama 223-8522, Japan}

\begin{abstract}
Circularly polarized photons have the Berry curvature in the semiclassical regime. 
Based on the kinetic equation for such chiral photons, we derive the (non)equilibrium 
expression of the photon current in the direction of the vorticity. We briefly discuss the 
relevance of this ``photonic chiral vortical effect" in pulsars and rotating massive stars and 
its possible realization in semiconductors. 
\end{abstract}
\maketitle

\section{Introduction}
Recently, the effect of the Berry curvature for photons has attracted great interest in optics 
and photonics. This effect, originating from the helical nature of circularly polarized photons, 
leads to remarkable topological transport phenomena. One canonical example of such 
phenomena is the quantum spin Hall effect of light \cite{Bliokh:2004, Onoda:2004zz}, which 
had also previously been known as the ``optical Magnus effect" \cite{Liberman:1992zz}.

In this paper, we develop a kinetic framework for right- and left-handed circularly polarized 
photons, similar to the one for spin-$1/2$ chiral fermions (known as the chiral kinetic theory) 
\cite{Son:2012wh, Stephanov:2012ki, Son:2012zy, Chen:2012ca}. Based on the kinetic 
equation, we derive a new type of topological transport phenomenon of photons in a 
rotation---the \emph{photonic chiral vortical effect} (CVE). This is the photon current along 
the direction of the vorticity. A similar CVE is known to appear in chiral matter that includes 
chiral fermions \cite{Vilenkin:1979ui, Kharzeev:2007tn, Son:2009tf, Landsteiner:2011cp} 
and has been intensively studied due to its possible relevance to quark matter in heavy ion 
collisions \cite{Kharzeev:2007tn, Son:2009tf} and neutrino matter in supernovae \cite{Yamamoto:2015gzz}. 
We argue that the photonic CVE provides a hitherto neglected contribution to the photon 
emission from pulsars and rotating massive stars. We also discuss a possible realization of 
this effect with a \emph{nonzero} chemical potential in semiconductors. We emphasize that 
our work enlarges the chiral transport phenomena so far limited to (nearly) massless chiral 
fermions to a drastically wider area of physical systems involving massless photons.

\section{Quantum mechanics for photons}

\subsection{Wave equation}
\label{sec:wave}
We first briefly recapitulate the wave equation for photons 
(see, e.g., Ref.~\cite{BialynickiBirula:1997am} for a review). 
To keep quantum mechanical and relativistic nature apparent, we will write explicitly 
$\hbar$ and $c$ in this section.


The wave function of photons must satisfy the following requirements:
\begin{itemize}
\item{It is linear, such that it can be superposed to have interference effects.}
\item{The coefficients are constants unrelated to the specific motion of photons:
$\hbar$ and/or $c$.}
\item{It satisfies the relation $\omega = c k$, with $\omega$ the frequency and 
$k \equiv |{\bm k}|$ the wave number.}
\end{itemize}
The first two conditions account for the wave nature of photons similar to the Schr\"odinger 
equation for electrons, and the third condition ensures the dispersion relation of massless 
photons or electromagnetic waves.

Before proceeding further, let us recall the basic properties of the two polarizations of 
photons (which we denote as ${\bm e}_{1}$ and ${\bm e}_{2}$) propagating in the 
direction specified by the wave vector ${\bm k}$:
\begin{gather}
{\bm e}_1 \cdot {\bm e}_2 = 0, \qquad {\bm k} \cdot {\bm e}_{1} = {\bm k} \cdot {\bm e}_{2} = 0, 
\\
{\bm k} \times {\bm e}_1 = k {\bm e}_2, \qquad {\bm k} \times {\bm e}_2 = -k {\bm e}_1,
\end{gather}
or equivalently, 
\begin{gather}
\label{e1}
k {\bm e}_{\pm} = \pm i {\bm k} \times {\bm e}_{\pm}, \\
\label{e2}
{\bm k} \cdot {\bm e}_{\pm} = 0,
\end{gather}
where we defined ${\bm e}_{\pm} \equiv {\bm e}_1 \pm i {\bm e}_2$. 

Now we take the wave functions of right- and left-handed photons, ${\bm \psi}_{\pm}(t, {\bm x})$, 
to be proportional to the complex polarizations ${\bm e}_{\pm}$. (Note that ${\bm \psi}_{\pm}$ 
are three-component wave functions.)
Recalling the properties (\ref{e1}) and (\ref{e2}), it turns out that the following equations satisfy 
all the requirements for the wave equation above:
\begin{gather}
\label{wave-eq1}
i \d_t {\bm \psi}_{\pm} = \pm c {\bm \nabla} \times {\bm \psi}_{\pm}\,, \\
\label{wave-eq2}
{\bm \nabla} \cdot {\bm \psi}_{\pm} = 0,
\end{gather}
for right- and left-handed photons, respectively. 
Equation (\ref{wave-eq1}) can be rewritten in the form of the Schr\"odinger-type equation as
\beq
i \hbar \d_t {\bm \psi}_{\pm} = \pm c \Bigl( {\bm S} \cdot  \frac{\hbar}{i} {\bm \nabla} \Bigr) {\bm \psi}_{\pm}\,,
\eeq
where $(S_i)_{jk} = -i \epsilon_{ijk}$ with the indices $i,j,k$ running over $1,2,3$. 

The corresponding Hamiltonian of chiral photons is thus given by
\beq
\label{H}
H = \pm c{\bm S} \cdot {\bm p},
\eeq
where ${\bm p}$ is the momentum. Note that $3 \times 3$ matrix $S_i$ ($i=1,2,3$) satisfies the 
commutation relations, $[S_i, S_j] = i \epsilon_{ijk} S_k$. This should be contrasted with the 
Hamiltonian of chiral fermions, $H = \pm c {\bm \sigma} \cdot {\bm p}$, where $\sigma_i$ is the 
$2 \times 2$ Pauli matrix that satisfies the commutation relations, 
$[\sigma_i, \sigma_j] = 2i \epsilon_{ijk} \sigma_k$.

In the medium where Lorentz symmetry is explicitly broken, the Hamiltonian (\ref{H}) is modified 
to 
\beq
H = \pm v {\bm S} \cdot {\bm p}, 
\eeq
where $v = 1/\sqrt{\epsilon \mu}$ is the velocity in medium with $\epsilon$ and $\mu$ being 
permittivity and permeability, respectively.

\subsection{Path integral formulation}
\label{sec:path}
In the following, we use the natural units $\hbar = c = 1$ for simplicity.
In order to derive the semiclassical theory for chiral photons with Berry curvature effects,
we consider the path integral formulation for the Hamiltonian (\ref{H}). 
Let us start with the path integral for right-handed photons ${\bm \psi}_{+}$, 
\beq
\label{path}
Z= \int {\cal D}x {\cal D}p {\cal P} e^{iI}\,, \qquad
I= \int{\rm d}t\,({\bm p} \cdot \dot {\bm x} - {\bm S} \cdot {\bm p}) \,,
\eeq
where ${\cal P}$ denotes the path-ordered product of the matrices $\exp(-i {\bm S} \cdot {\bm p} \Delta t)$ 
over the path in the phase space. The following argument can be similarly applied for left-handed 
photons ${\bm \psi}_{-}$ as well.

The eigenvalues of the $3 \times 3$ matrix ${\bm S} \cdot {\bm p}$ are found to be $\pm |\bm p|$ 
and $0$. One can diagonalize this matrix using a unitary matrix $V_{\bm p}$, such that
\begin{equation}
V_{\bm p}^{\dag} {\bm S} \cdot {\bm p} V_{\bm p} = 
\begin{pmatrix}
|{\bm p}| & 0 & 0 \\
0 & -|{\bm p}| & 0 \\
0 & 0 & 0
\end{pmatrix} 
\equiv |\bm p| \lambda_3 \,,
\end{equation}
where $\lambda_3 ={\rm diag}(1,-1,0)$ is one of the ${\rm SU}(3)$ generators. 
The eigenstate of the eigenvalue $-|{\bm p}|$ has negative energy and should be regarded 
as unphysical for photons. This is to be contrasted with the case of chiral fermions, where 
negative energy states, corresponding to antiparticles, are possible.
The eigenstate of the eigenvalue $0$ is given by $\hat {\bm p} \equiv {\bm p}/|{\bm p}|$ 
(multiplied by any nonzero proportionality constant) and is longitudinal with respect to ${\bm p}$. 
Due to the additional constraint (\ref{wave-eq2}), this state is forbidden to appear and is 
unphysical as well; hence, ${\bm \psi}_+$ has only one physical eigenstate with the eigenvalue 
$|{\bm p}|$, corresponding to the positive helicity state $h = +1$.

Following Ref.~\cite{Stephanov:2012ki}, one can rewrite the path integral (\ref{path}) by 
inserting $1 = V_{\bm p}V_{\bm p}^{\dag}$ between the exponential factors, so that the matrix 
in the exponential factor is diagonalized at each point of the trajectory as follows:
\begin{align}
\label{factor}
\cdots \exp(-i {\bm S} \cdot {\bm p}_2 \Delta t)\exp(-i {\bm S} \cdot {\bm p}_1 \Delta t) \cdots 
&= \cdots V_{{\bm p}_2} V_{{\bm p}_2}^{\dag} \exp(-i {\bm S} \cdot {\bm p}_2 \Delta t) V_{{\bm p}_2} V_{{\bm p}_2}^{\dag} 
\nonumber \\
& \qquad \times V_{{\bm p}_1} V_{{\bm p}_1}^{\dag} \exp(-i {\bm S} \cdot {\bm p}_1 \Delta t) V_{{\bm p}_1} V_{{\bm p}_1}^{\dag} \cdots 
\nonumber \\
& = \cdots V_{{\bm p}_2} \exp(-i |{\bm p}_2| \lambda_3 \Delta t) V_{{\bm p}_2}^{\dag} 
\nonumber \\
& \qquad \times 
V_{{\bm p}_1} \exp(-i |{\bm p}_1| \lambda_3 \Delta t) V_{{\bm p}_1}^{\dag} \cdots \,.
\end{align}
Taking $\Delta {\bm p} \equiv {\bm p}_2 - {\bm p}_1$ to be sufficiently small, the factor 
$V_{{\bm p}_2}^{\dag} V_{{\bm p}_1}$ between the two exponential factors in Eq.~(\ref{factor}) 
can be expressed as
\beq
\label{V}
V_{{\bm p}_2}^{\dag} V_{{\bm p}_1} \approx \exp(-i \hat {\bm a}_{\bm p} \cdot \Delta {\bm p})
= \exp(-i \hat {\bm a}_{\bm p} \cdot \dot {\bm p} \Delta t),
\eeq
where $\hat {\bm a}_{\bm p} \equiv i V_{\bm p}^{\dag} {\bm \nabla}_{\bm p} V_{\bm p}$.

We now take the semiclassical limit where off-diagonal components of $\hat {\bm a}_{\bm p}$ 
are negligible. (We will discuss the validity of this approximation later.) Focusing on the positive 
energy state, we arrive at the semiclassical action for right-handed photons,
\beq
\label{I}
I= \int {\rm d}t\,({\bm p} \cdot \dot {\bm x} - {\bm a}_{\bm p} \cdot \dot {\bm p} - \epsilon_{\bm p}),
\eeq 
where $\epsilon_{\bm p} = |{\bm p}|$ is the energy dispersion (in the vacuum) and 
${\bm a}_{\bm p} \equiv [{\hat {\bm a}}_{\bm p}]_{11}$ is the gauge field in momentum space,
called the Berry connection. The corresponding field strength, called the Berry curvature, is 
defined as ${\bm \Omega}_{\bm p} \equiv {\bm \nabla}_{\bm p} \times {\bm a}_{\bm p} $. 
Similarly, one can obtain the semiclassical action for left-handed photons (or negative helicity 
state $h=-1$) by repeating the similar argument for ${\bm \psi}_-$.

From the definition of ${\bm a}_{\bm p}$ above, one finds that 
\beq
\label{curvature}
{\bm \Omega}_{\bm p} = \pm \frac{\hat {\bm p}}{|\bm p|^2}\,,
\eeq
for right- and left-handed photons with $h = \pm 1$, respectively. 
This is the fictitious magnetic field of the magnetic monopole (in momentum space) with the 
monopole charge,
\beq
k = \frac{1}{4\pi} \int {\bm \Omega}_{\bm p} \cdot {\rm d}{\bm S} = \pm 1\,,
\eeq
where the area integral is taken over the surface of the sphere with radius $|{\bm p}|$. 
Note that the Berry curvature of chiral photons in Eq.~(\ref{curvature}) is twice larger than 
that of chiral fermions in Refs.~\cite{Son:2012wh, Son:2012zy, Stephanov:2012ki}.

Let us discuss the applicability of the semiclassical description for photons above. 
For the off-diagonal components of $\hat {\bm a}_{\bm p} \cdot \dot {\bm p}$ to be negligible 
to obtain Eq.~(\ref{I}) from Eq.~(\ref{V}), they must be much smaller than the energy gap 
$2|{\bm p}|$ between the two eigenstates with the eigenvalues $\pm |{\bm p}|$. As 
$|\hat {\bm a}_{\bm p}| \sim 1/|{\bm p}|$, this condition amounts to 
\beq
\label{condition}
|\dot {\bm p}| \ll |{\bm p}|^2,
\eeq
meaning that ${\bm p}$ must be sufficiently away from the level crossing point ${\bm p} = {\bm 0}$.

\subsection{Semiclassical equations of motion}
\label{sec:eom}
Let us look into the consequences of the Berry curvature corrections for chiral photons.
From the action (\ref{I}), one obtains the equation of motion for $\dot {\bm x}$,
\beq
\label{xdot}
\dot {\bm x} &= \hat {\bm p} + \dot {\bm p} \times {\bm \Omega}_{\bm p}.
\eeq
The second term on the right-hand side of Eq.~(\ref{xdot}) is the ``Lorentz force" in 
momentum space, originally known as the optical Magnus effect \cite{Liberman:1992zz}. 
This term has been found to induce the spin Hall effect of light \cite{Bliokh:2004, Onoda:2004zz}:
the trajectory of the circularly polarized photon is shifted perpendicularly to the direction of 
$\dot {\bm p}$. For example, in an inhomogeneous medium with coordinate-dependent 
permittivity $\epsilon(\bm x)$, the equation of motion is $\dot {\bm p} = - ({\bm \nabla}v)|{\bm p}|$, 
where $v= 1/\sqrt{\epsilon}$. In this case, the shift of the trajectory is perpendicular to the 
direction of ${\bm \nabla} \epsilon$ \cite{Liberman:1992zz, Bliokh:2004, Onoda:2004zz}.

\section{Photonic chiral vortical effect}
\label{sec:PCVE}
As we will discuss from now on, the Berry curvature correction of photons also leads to a new 
type of topological transport phenomenon---the photonic CVE.

Let us consider the response of a system with right- or left-handed photons to a global rotation 
or a local vorticity ${\bm \omega} = \frac{1}{2}{\bm \nabla} \times {\bm v}$, where ${\bm v}$ is 
the local fluid velocity.%
\footnote{In the case of a system in a global rotation ${\bm \omega}$, the size of the system
of interest is to be understood as satisfying $r < 1/|\bm \omega|$. Otherwise, the velocity of 
the boundary exceeds the speed of light, leading to unphysical results \cite{Vilenkin:1979ui}
(see also Ref.~\cite{Davies:1996ks}).}
For this purpose, let us go to the comoving frame rotating with angular velocity ${\bm \omega}$ 
with respect to the laboratory frame, similarly to Ref.~\cite{Stephanov:2012ki}. In this frame, 
photons experience the noninertial Coriolis force to the linear order in ${\bm \omega}$. 
We will be interested in sufficiently small $|{\bm \omega}|$ so that the centrifugal force of order 
$O({\bm \omega}^2)$ is negligible. Then the equation of motion for $\dot {\bm p}$ 
is given by
\beq
\label{pdot}
\dot {\bm p} = 2|{\bm p}| \dot {\bm x} \times {\bm \omega} + O({\bm \omega}^2),
\eeq
where we assumed a homogeneous medium for simplicity. The right-hand side of 
Eq.~(\ref{pdot}) can be understood as a relativistic generalization of the Coriolis force 
$2 m \dot {\bm x} \times {\bm \omega}$ for a nonrelativistic particle with mass $m$.
Substituting Eq.~(\ref{pdot}) into Eq.~(\ref{xdot}), we have
\beq
\label{xdot_G}
\sqrt{G} \dot {\bm x} 
= \hat {\bm p} + 2{\bm \omega} |{\bm p}|(\hat {\bm p} \cdot {\bm \Omega}_{\bm p}),
\eeq
where $G = (1 + 2 |{\bm p}| {\bm \omega} \cdot {\bm \Omega}_{\bm p})^2$ is the determinant 
of the $6 \times 6$ matrix of the coefficients in Eqs.~(\ref{xdot}) and (\ref{pdot}) for 
$\dot {\bm x}$ and $\dot {\bm p}$. 

Using the distribution function of right- or left-handed photons in the phase space, $n_{\bm p}$, 
the photon current density is given by
\beq
{\bm j} = \int \frac{{\rm d}^3{\bm p}}{(2\pi)^3} \sqrt{G} \dot {\bm x} n_{\bm p}\,, 
\eeq
where we took into account the fact that the invariant phase space measure becomes
$\sqrt{G}{\rm d}^3{\bm x} {\rm d}^3{\bm p}/(2\pi)^3$ instead of 
${\rm d}^3{\bm x} {\rm d}^3{\bm p}/(2\pi)^3$ due to the modification in Eq.~(\ref{xdot_G}).
This modification is similar to that of chiral fermions in a magnetic field 
\cite{Son:2012wh, Son:2012zy, Stephanov:2012ki} (see also Refs.~\cite{Xiao:2005, Duval:2005}).

From the Berry-curvature corrections in Eq.~(\ref{xdot_G}), one finds the photon current 
proportional to the vorticity,
\beq
\label{PCVE}
{\bm j}_{\rm CVE} = 2{\bm \omega}\int \frac{{\rm d}^3{\bm p}}{(2\pi)^3}
|{\bm p}|(\hat {\bm p} \cdot {\bm \Omega}_{\bm p}) n_{\bm p} \,.
\eeq
This is the nonequilibrium expression of the photonic CVE.
Note that, although the expression itself seems the same as the CVE for chiral fermions in 
Ref.~\cite{Stephanov:2012ki}, this is indeed different from the latter: the Berry curvature 
${\bm \Omega}_{\bm p}$ in Eq.~(\ref{curvature}) is twice larger than the one in 
Ref.~\cite{Stephanov:2012ki} and $n_{\bm p}$ is the bosonic distribution function unlike 
the fermionic one in Ref.~\cite{Stephanov:2012ki}.

These differences can be clearly seen in the thermal equilibrium state where $n_{\bm p}$ takes 
the Bose-Einstein distribution characterized by temperature $T$ and chemical potential $\mu$,%
\footnote{Usually the chemical potential of photons is vanishing, $\mu = 0$, because the number 
of photons can vary without any constraint. However, this is not always true when photons are in 
chemical equilibrium with the excitations of matter with a nonzero chemical potential; e.g., 
chemical equilibrium between photons and electron-hole pairs in a light-emitting diode can lead to 
a photon medium with $\mu \neq 0$ \cite{Wurfel}; see also Sec.~\ref{sec:discussion}.
Here we consider the most generic case with nonzero $T$ and $\mu$.}
\beq
n_{\bm p} = \frac{1}{{\rm e}^{\beta (\epsilon_{\bm p} - \mu)}-1}\,,
\eeq
with $\beta \equiv 1/T$. In this case, the photonic CVE becomes
\beq
\label{PCVE1}
{\bm j}_{\rm CVE}^{\pm} = \pm{\bm \omega} \int_0^{\infty} \frac{{\rm d}p}{\pi^2} \frac{p}{{\rm e}^{\beta(p-\mu)}-1}
= \pm \frac{1}{\pi^2} F(2, -\beta \mu) T^2 {\bm \omega}\,,
\eeq
for right- and left-handed photons, respectively, where $p \equiv |{\bm p}|$ and
\beq
F(s, \alpha) \equiv \frac{1}{\Gamma(s)}\int_0^{\infty} \frac{x^{s-1}}{e^{x+\alpha}-1} {\rm d}x
= \sum_{n=1}^{\infty} \frac{{\rm e}^{-n \alpha}}{n^s}
\eeq
is the Bose-Einstein integral with $\Gamma(s)$ the gamma function. Apparently, the transport 
coefficient in Eq.~(\ref{PCVE}) is different from that for chiral fermions in 
Ref.~\cite{Stephanov:2012ki}. This is one of the main results in this paper. 

In particular, when $\mu=0$, the expression of the photonic CVE is further simplified by using 
$F(2,0) = \pi^2/6$ as
\beq
\label{PCVE2}
{\bm j}_{\rm CVE}^{\pm} = \pm \frac{T^2}{6} {\bm \omega}\,.
\eeq
Incidentally, the transport coefficient in this case is the same as that for chiral matter with a 
single chiral fermion. This is a consequence of two modifications compared with chiral fermions, 
which cancel with each other: one is the fact that the helicity of photons, $h=\pm1$, is twice larger 
than the helicity of chiral fermions, $h=\pm 1/2$ (and so is the Berry curvature), and the other is 
that the contribution of antiparticles is absent for photons.

Notice that the coefficients of the photonic CVE have the opposite signs for thermalized 
right- and left-handed photons. Hence, in a system with \emph{both} right- and left-handed 
photons in a rotation, they tend to move in the opposite direction. The corresponding axial 
current at finite $T$ and $\mu$ is expressed as
\beq
\label{PCSE}
{\bm j}^{\rm A} \equiv {\bm j}^{+} - {\bm j}^{-} = \frac{2}{\pi^2} F(2, -\beta \mu) T^2 {\bm \omega}\,.
\eeq
As a result, right- and left-handed photons are separated along the rotation. 
This is the \emph{photonic chiral separation effect}.

One might think that the argument leading to Eq.~(\ref{PCVE1}) above would not be completely 
justified because the integration over momentum space in Eq.~(\ref{PCVE1}) includes the 
singular point ${\bm p} = {\bm 0}$, where the semiclassical description for photons breaks down. 
In fact, the condition~(\ref{condition}), together with the equations of motion (\ref{xdot}) and 
(\ref{pdot}), requires that $|{\bm \omega}| \ll |{\bm p}|$. However, one can show that the 
contribution around the singular point with the region $|{\bm p}| \leq \Delta$ (with $\Delta$ 
satisfying $|{\bm \omega}| \ll \Delta \ll T$) to the integral in Eq.~(\ref{PCVE1}) is vanishingly small; 
when $\mu \neq 0$, we have
\beq
\int_0^{\Delta}\frac{{\rm d}p}{\pi^2} \frac{p}{{\rm e}^{\beta (p-\mu)}-1} \sim \Delta^2 \ll T^2\,,
\eeq
and when $\mu = 0$,
\beq
\int_0^{\Delta}\frac{{\rm d}p}{\pi^2} \frac{p}{{\rm e}^{\beta p}-1} \sim T \Delta \ll T^2\,,
\eeq
where we used ${\rm e}^{\beta p} \simeq 1 + \beta p$ for $p \ll T$. Hence, the support of the 
integrand in Eq.~(\ref{PCVE1}) comes from the region $p > \Delta$, where the semiclassical 
treatment is valid.

\section{Discussions}
\label{sec:discussion}
In this paper, we developed a kinetic description for right- and left-handed circularly polarized
photons. Using the kinetic equation, we derived the expression of the (non)equilibrium photon 
number current along the direction of a vorticity. The nonequilibrium and equilibrium photonic 
chiral vortical effects are given by Eqs.~(\ref{PCVE}) and (\ref{PCVE1}), respectively. 

Among others, the photonic CVE may provide a novel contribution to the photon emission from 
pulsars and rotating massive stars.%
\footnote{We note that an attempt to associate the usual \emph{fermionic} CVE with 
astrophysical jets was made in Ref.~\cite{Flachi:2017vlp}.}
This effect is remarkable in that photons emitted along the rotational axis of a star have a
dependence on the circular polarizations: only right-handed photons are emitted from one of 
the poles while only left-handed ones from the other. As the photonic CVE becomes larger as 
the temperature and angular velocity increase [see Eq.~(\ref{PCVE2})], we expect this effect 
to be most significant in the accretion-powered millisecond pulsars. A more detailed analysis
in this direction will be reported elsewhere.
One can extend the conventional framework of radiation hydrodynamics for matter-photon 
coupled astrophysical systems to include this effect by using the chiral kinetic theory for 
photons considered in this paper \cite{Yamamoto}. 

The photonic CVE with a \emph{nonzero} chemical potential may also be realized in table-top 
experiments. In fact, thermalized photon media with finite temperature and chemical potential 
can be produced in semiconductors (e.g., light-emitting diodes) in an external electric field 
due to the chemical reactions of electrons and holes with photons \cite{Wurfel}.%
\footnote{This situation is somewhat similar to left-handed neutrinos at the core of supernovae; 
even weakly interacting neutrinos can be thermalized and have nonzero chemical potential 
due to the weak equilibrium with nucleons and electrons in dense nuclear matter, where the CVE 
for neutrinos is relevant to the evolution of core-collapse supernovae \cite{Yamamoto:2015gzz}.}
Then, such a system in a rotation should exhibit the photonic chiral separation effect: 
right- and left-handed circularly polarized photons are separated along the rotation. 

As the CVE appears not only for the spin-1/2 chiral fermions, but also for spin-1 chiral 
photons as we argued in this paper, one expects the same is also the case for higher-spin 
massless chiral particles (such as chiral gravitons). It would also be interesting to study 
possible new collective modes induced by the photonic CVE 
(see Refs.~\cite{Jiang:2015cva, Yamamoto:2015ria, Chernodub:2015gxa, Abbasi:2015saa, 
Kalaydzhyan:2016dyr, Abbasi:2016rds} for the case of chiral fermions).
Finally, one may be able to derive the chiral kinetic theory for photons from the microscopic 
quantum field theory in a way similar to the one for chiral fermions 
\cite{Son:2012zy, Manuel:2014dza, Hidaka:2016yjf, Mueller:2017lzw}.

\section*{Acknowledgement}
The author thanks T.~Enoto for useful conversations. 
This work was supported by JSPS KAKENHI Grant No.~16K17703 and MEXT-Supported 
Program for the Strategic Research Foundation at Private Universities, ``Topological Science" 
(Grant No.~S1511006).

\emph{Note added.}---Recently, we learned that A.~Avkhadiev and A.~V.~Sadofyev 
\cite{Avkhadiev:2017fxj} and V.~A.~Zyuzin \cite{Zyuzin} have independently studied the chiral 
vortical effect for bosons by different approaches. We note, however, that our derivation and 
results, based on the kinetic theory with Berry-curvature corrections, are more generic than 
those of Refs.~\cite{Avkhadiev:2017fxj,Zyuzin} in that the former is applicable to the 
\emph{nonequilibrium} state (and equilibrium state with finite $T$ and $\mu$), while the latter is 
limited to the \emph{equilibrium} state without $\mu$.

\end{document}